\documentclass[%
reprint,
 amsmath,amssymb,
 aps,
]{revtex4-1}

\usepackage{graphicx}
\usepackage{dcolumn}
\usepackage{bm}
\usepackage{color}
\usepackage{lineno}

\begin{document}

\preprint{APS/123-QED}

\title{Challenges of interpreting dielectric dilatometry for the study of pressure densification}

\author{Alejandro Sanz}\email{asanz@ruc.dk}
\author{Jeppe C. Dyre}%
\author{Kristine Niss}
\affiliation{Glass and Time, IMFUFA, Department of Science and Environment, Roskilde University, Postbox 260, DK-4000 Roskilde, Denmark}

\date{\today}

\begin{abstract}
	
We report an experimental study documenting the challenge of employing dielectric dilatometry for the study of pressure densification in glass-forming materials. An influence of the dielectric cell geometry on the resulting capacitance of 5-poly-phenyl-ether upon vitrification under different thermobaric pathways is documented. The capacitive response is studied for two different multilayer capacitors: one with, in principle, fixed plate distance and one with Kapton$\textregistered$ spacers allowing for contraction/expansion. A combination of changes in the dielectric permittivity of the material and modifications of the capacitor geometry determines the final capacitance. We conclude that, in order to convert the measured capacitance to material density, it is of paramount importance to understand the geometry. The data presented do not make it possible to conclude on whether or not simple glass formers
such as 5-poly-phenyl-ether can be pressure densified, but our work highlights the challenge of utilizing dielectric spectroscopy to tackle this problem effectively.

\begin{description}
\item[PACS numbers]
64.70.kj, 64.70.pm, 62.50.-p, 77.22.-d.
\end{description}
\end{abstract}

\pacs{Valid PACS appear here}
\maketitle


\section{Introduction}

Density as a function of temperature and pressure is
among the most fundamental thermodynamic and structural parameters of
a glass-forming liquid, and the abrupt change in the temperature and pressure derivatives of density, the thermal expansion
coefficient and the compressibility, are a principal signature of the
glass transition \cite{Kristine4}. Low-molecular-weight glass-forming liquids are often used
in fundamental research on the glass transition \cite{Ediger2,Debenedetti,Qin,Nielsen,TinaNP}.
However, measuring the density in the glass transition region is
difficult for these liquids due to their low glass transition
temperatures.  Most techniques for measuring density rely on a
confining liquid and are limited by the temperature-pressure range
where it is in its liquid state, e.g. mercury which has a melting point
of 234.3~K at atmospheric pressure. This challenge has the consequence
that a fundamental result like density scaling to a
large extent is based on extrapolated density data when it comes to
molecular liquids \cite{Romanini,Grzybowski,Hansen}.

An alternative method for measuring density is dielectric dilatometry
\cite{Bauer,Kristine2}. This technique makes use of the fact that the
dielectric constant of a material depends on the number density
of molecules (and number density is proportional to mass density, as long as
no chemical reactions take place). The advantage of dielectric
dilatometry is that it can be measured with very high precision and
that it can be applied at low temperatures. The method has been used
in several studies of scanning dilatometry of polymer films where it
has the additional advantage of getting more precise with a thinner film
\cite{Bauer,Fukao,Yin,Inoue}. It has also been used for measuring the dynamic linear expansion coefficient where the high precision gained by this method is crucial for obtaining the resolution needed to measure in the linear range \cite{Bauer, Kristine2,Kristine3}. Recently, the method has been employed to study pressure densification of a molecular glass \cite{Casalini2}. While dielectric dilatometry has potential to be used more in the field, it also has some limitations. In this paper we investigate these limitations
with a focus on the challenges connected to study pressure densification.

Pressure densification refers to a situation where a glass formed by
cooling at elevated pressures and subsequently decompressed in the glassy
state, has a higher density than a glass of the same material formed by
cooling at low pressures. It is a physical phenomenon widely known in inorganic glasses, such as silica \cite{Tse,Valle,Rouxel}, that can be used for tuning its optical properties \cite{Masuno}. Roland and colleagues recently proposed a new approach for examining pressure densification from a more fundamental point of view. They suggested examining the problem under the framework of the isomorph theory \cite{Casalini2,Fragiadakis,Casalini2017}, which predicts the existence of isomorphs in the so-called R-simple systems \cite{Jeppe}.
Pressure densification is also a long-standing topic in the physics of polymeric materials that has been investigated both experimentally and theoretically, for instance in polymethylmethacrylate (PMMA) and in other amorphous polymers \cite{Kimmel,Destruel,Casalini2017}. Pressure densification has also been studied in polymeric systems by computer simulations \cite{Fragiadakis}.
The fact that PMMA can be pressure densified is somewhat surprising since this polymer conforms with isochronal superposition and density scaling \cite{Casalini2004}, and therefore it is likely to behave as a "Roskilde-simple system" (R-simple) according to the isomorph theory \cite{Gnan,Jeppe}.

The increase in density of glasses prepared under high pressure was found to vary from 1-2 \% in molecular glasses and polymers to more than 15 \% in inorganic materials \cite{Danilov,Kimmel,Rouxel}. Casalini and Roland proposed a metric for the degree of densification in terms of the relative changes in specific volume for vitrification at high and low pressures. This so-called pressure densification ratio, $\delta$, was estimated to be around 25 \% for densification pressures between 300 and 570 MPa in the molecular glass-former tetramethyl-tetraphenyl-siloxane (DC704) \cite{Casalini2}.     

The recent interest in pressure densification from a more fundamental
point of view is sparked by the prediction that R-simple liquids cannot be pressure densified
\cite{Gnan2,Fragiadakis}. Isomorph theory predicts that
all glasses formed along the same glass-transition isochrone (for
example formed with the same cooling rate at different pressures) will have
the same microscopic structure, except for the average distance
between the molecules which will scale with the qubic root of the
inverse density. If the pressure is released on a glass formed at high
pressure, then the average molecular distance will increase, and the
glass will attain the same density as the glass formed at lower
pressures. This has been supported by simulations but some experiments contradict this prediction \cite{Danilov,Casalini2,Casalini2017}.

\begin{figure}
	\includegraphics[trim = 60mm 47mm 15mm 55mm, clip, width=0.7\textwidth]{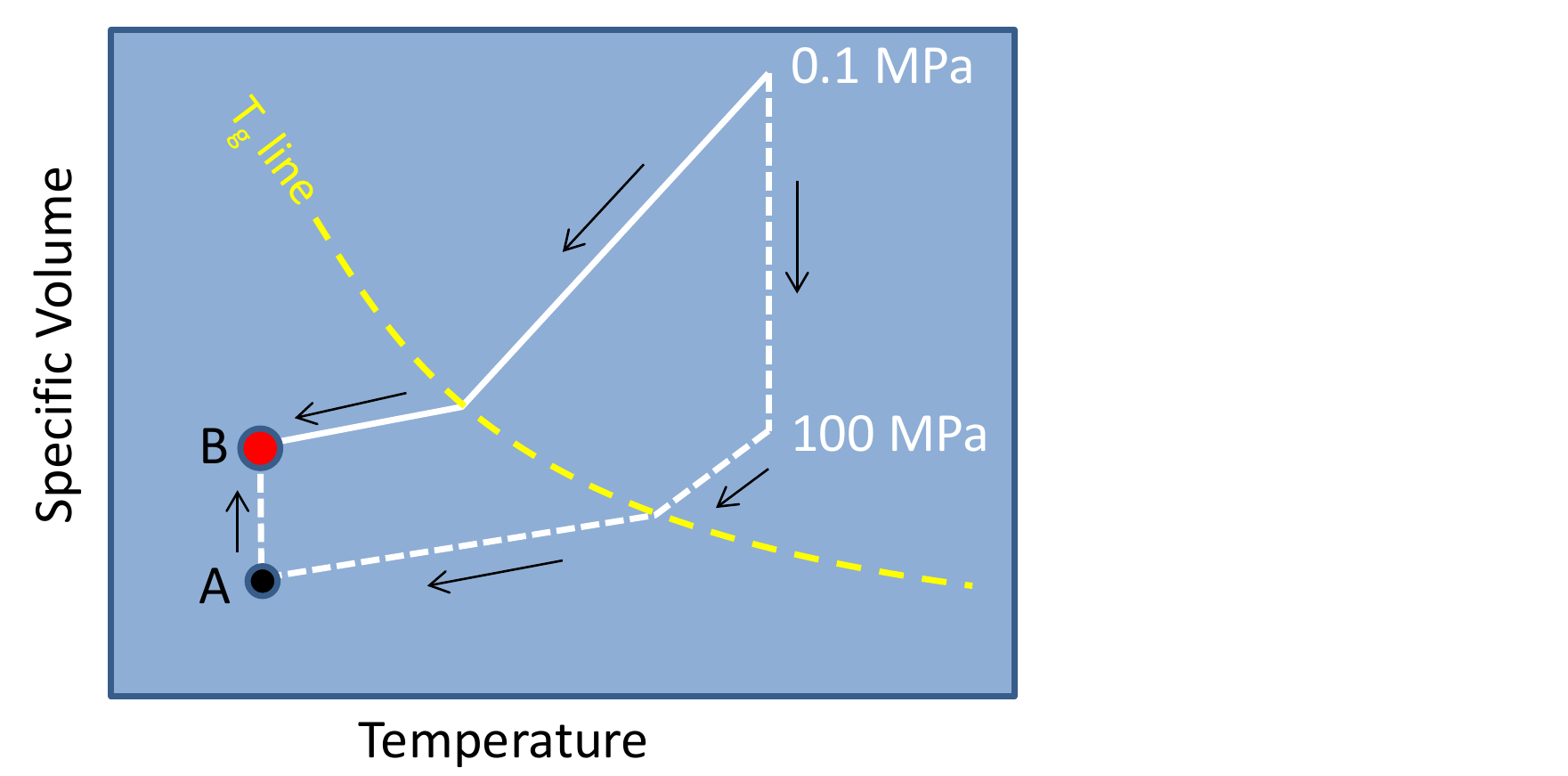}
	\caption{Schematic phase diagram showing the dependence of specific volume
		with temperature at 0.1 and 100 MPa upon cooling through the
		glass transition. Solid and dashed white lines describe the
		thermodynamic routes for reaching the state point B explored
		in this work. The dashed yellow line corresponds to the $T_{g}$
		line.}
	\label{fig:Fig1}
\end{figure}

In this study, we have investigated experimentally two
temperature-pressure protocols, using two different dielectric
cells in order to highlight the role of macroscopic geometry in these
experiments. A schematic description of the phase diagram explored in this work is illustrated in Fig. \ref{fig:Fig1}.  

In the following section we present the fundamental principle of
dielectric dilatometry. In section III we present the experimental method. Section IV presents the results which are analyzed and discussed in section V. 

\section{Background: Dielectric dilatometry}

The fundamental assumption that needs to be fulfilled in order to
use dielectric dilatometry is that the high-frequency dielectric permittivity, $\varepsilon_\infty$, only
depends on density in the part
of pressure-temperature diagram where it is measured. On a quantitative
level, the high-frequency dielectric permittivity is connected to molecular properties
using the Clausius-Mossotti (CM) relation \cite{Bottcher}:
\begin{equation}\label{Eq1}
\frac{\epsilon_{\infty}-1}{\epsilon_{\infty}+2}=\frac{n}{3\epsilon_0}\alpha_0,
\end{equation}
where $\alpha_0$ is the polarizability of a single molecule, $n$ is
the number density of molecules and $\epsilon_0$ is the vacuum
permittivity. It is seen from this expression that $\alpha_0$ needs to
be temperature and pressure independent in order for the method to be
viable for determining $n$. In contrast, the polarization due to rotation of permanent dipoles is
highly pressure and temperature dependent in glass-forming liquids. In
fact, dielectric spectroscopy is among the most commonly applied
techniques to investigate dynamics of glass-formers \cite{Kremer,TinaNP}. However, the
rotational contribution is frozen out as the glass transition is
approached, and if measurements are performed at high frequencies close
to the glass transition, the rotational contribution can in many cases
be assumed to be negligible. Besides rotation of permanent dipoles,
the other contribution to $\alpha_0$ is the atomic polarization
due to deformation of the electron cloud. The atomic polarization takes place instantaneously and it can be assumed to be constant as long as temperatures are much lower than
the electronic binding energies. Given the local character of the induced atomic polarization and the modest pressures used in the present study, atomic polarization can be considered pressure-independent, as well.

The CM-relation is derived as a mean field approximation, and is
not expected to be exact. Deviations from CM may influence the
absolutes values found from the method, but will not affect
temperature, pressure or time dependence found of the thermal
expansion coefficient or the bulk modulus (see Ref. \cite{Kristine2} for
details).

\begin{figure*}
	\includegraphics[trim = 20mm 30mm 20mm 5mm, clip, width=0.8\textwidth]{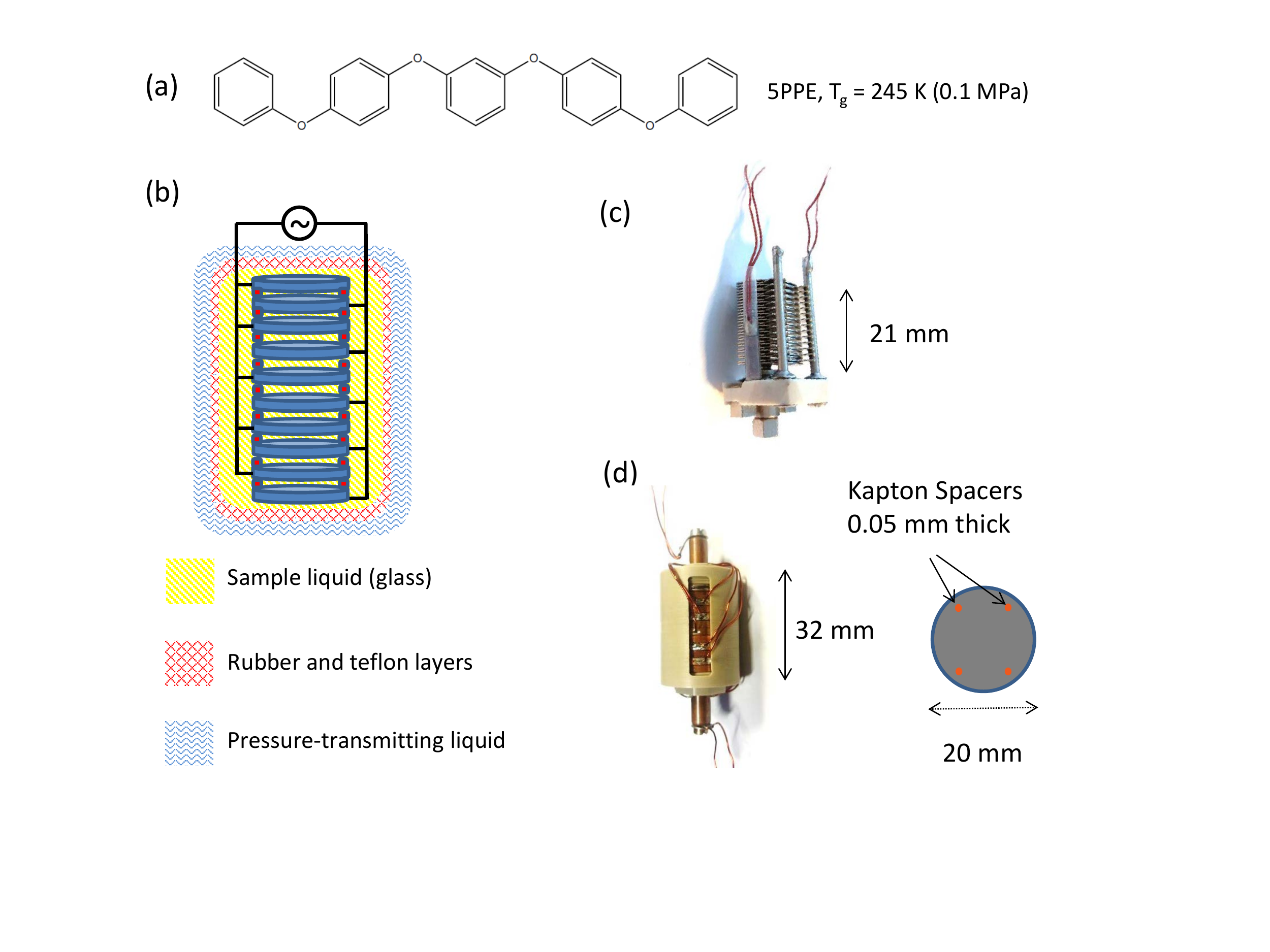}
	\caption{(a) Chemical structure of 5-poly-phenyl-ether and
		value of the glass transition temperature at atmospheric
		pressure. (b) Drawing of the experimental setup for
		measuring the frequency-dependent capacitance as a function
		of pressure and temperature. The sample liquid is sandwiched
		between metallic plates and the dielectric
		capacitor is isolated from the pressure-transmitting medium
		by successive layers of rubber and teflon tape. Red dots correspond to spacers which are used in the home-built capacitor shown in panel (d). Dimensions are not to scale for the
		sake of clarity. (c) Photograph of the air capacitor used in the present study. Results obtained from this
		dielectric cell are displayed in Fig. \ref{fig:Fig3} and \ref{fig:Fig4}. (d)
		Photograph of the multilayer capacitor where the spacing
		between the electrodes is controlled by 4 Kapton$\textregistered$ spacers
		with 0.05 mm thickness. Results obtained from this
		dielectric cell are displayed in Fig. \ref{fig:Fig5}.}
	\label{fig:Fig2}
\end{figure*} 

Given the connection between density and dielectric constant, the next
step is to connect the measured capacitance with the dielectric
constant. In the simple case with a capacitor completely filled with sample liquid one has 
\begin{equation}\label{Eq2}
C_m=\varepsilon_\infty C_g,
\end{equation}
where \textit{$C_g$} is the empty capacitance which is determined by
the geometry of the capacitor. For a parallel plate geometry it is
given by $C_g=\varepsilon_0 A/l$, with $l$ the distance between the
plates and $A$ the area.

The measured change in capacitance, following an applied temperature
or pressure variation, is a combined effect of the change in dielectric constant and the change in
the geometric capacitance
\begin{equation}\label{Eq3}
\textrm{d}C_m=\varepsilon_\infty \textrm{d}C_g+C_g\textrm{d}\varepsilon_\infty.
\end{equation}
The use of dielectric dilatometry consequently relies on assumptions
regarding how the geometrical capacitance responds to pressure and
temperature changes. What the adequate assumptions are, depend on the design and materials of the capacitor.

In the studies of polymeric films by dielectric dilatometry, the
standard construction is one where the electrodes of the capacitor are
directly mounted on the sample, such that the distance between the
electrodes is determined by the thickness of the polymeric film \cite{Kremer}. In this case, both
terms in Eq. (\ref{Eq3}) contribute and they are both proportional to the density. In our previous studies of molecular
liquids, a similar assumption was used \cite{Kristine2,Kristine3}.
The electrodes were separated by soft spacers which
made it possible to fill the capacitor with the sample in its fluid
state at room temperature. In the glass transition region, where the
measurements were performed, it was assumed that the sample determined
the electrode distance. This assumption is based on the consideration
that a glass (or a liquid close to the glass-transition) is unable
to flow in the small gap between the electrodes, and that the
stiffness of the glassy (or almost glassy) sample outmatches that of
the soft spacers (see Ref. \cite{Kristine2} for details).

In the studies of pressure densification in Ref. \cite{Casalini2}, the capacitor used
was an air capacitor of the kind similar to those used in old radios. In this case, there are no spacers
between the electrodes, and it is a metallic scaffold which keeps the
distance between electrodes. The authors therefore assumed that the
geometric capacitance was unchanged during the experiment and that the
measured capacitance was only affected by the change in dielectric
constant, determined by the number of molecules in the fixed sample
volume. This assumption implies that the sample is able to
flow in and out of the capacitor. However, when working with a sample
below the glass transition (as was the case in the crucial part of the
experiment) this assumption is questionable. A glass or even a highly viscous liquid does not flow unless
waiting times several orders of magnitude longer than the alpha
relaxation time are inferred \cite{Kristine2} (depending on the capacitor geometry).

\section{EXPERIMENTAL DETAILS AND MEASURING PRINCIPLE}

To investigate the role of capacitor geometry and boundary conditions
on pressure densification results, we have studied the same sample with
different protocols using two different capacitors. The sample studied is
the van der Waals liquid 5-poly-phenyl-ether (5PPE). The sample, with
chemical structure shown in Fig. \ref{fig:Fig2}(a), was purchased from
Santolubes and used as received. Due to its relatively high glass transition temperature
($T_g$) at atmospheric pressure, 245 K, 5PPE suits the technical
specifications of our high-pressure dielectric setup. Furthermore, 5PPE fulfils several prerequisites to be considered an R-simple liquid \cite{Lisa,Lisa2,Hansen,Hansen2}, just like
tetramethyl-tetraphenyl-siloxane studied in
Ref. \cite{Casalini2}. In 5PPE we thus expect the existence of isomorphs, that is, lines in the phase diagram along which several structural, dynamic and thermodynamic properties are invariant \cite{Lisa,Jeppe,Lisa2,Hansen,Hansen2}. Isomorphs are also isochrones, which are curves in the phase diagram with constant relaxation time.

Two different multilayer capacitors were used in the study; an air
capacitor and a home-built capacitor with Kapton$\textregistered$ spacers. Figure
\ref{fig:Fig2}(c) shows a picture of the air
capacitor. It is similar to that used in Ref. \cite{Casalini2} and has
no spacer material, because the electrode plates are kept apart by
the metallic scaffold. The distance between the electrode plates is in
the order of half a millimeter and the empty capacitance is 85 pF when measured at ambient
conditions. Figure \ref{fig:Fig2}(d) shows the home-built multilayer
capacitor. The separation between the metallic electrodes is
maintained by four spacers made of Kapton$\textregistered$ (50 $\mu$m
thick). The spacers determine the distance when the capacitor is empty
or filled with a low viscosity liquid, while the sample itself will
determine the distance between the plates for viscosities high enough
that the sample cannot flow \cite{Kristine2}. Twelve plates are stacked on top of
each other forming eleven capacitors in parallel giving a total empty
capacitance at ambient conditions of approximately 560 pF.

In both cases, once the dielectric capacitor is filled with the liquid under study at
room temperature, it is immersed in a rubber container full of the
same sample liquid. This reservoir makes it possible for extra liquid to flow
into the capacitor as the liquid contracts upon cooling, a flow that, notably, is only possible at low-viscosity conditions.
Successive rubber and teflon layers shield the sample from the
pressure-transmitting medium. This configuration is illustrated in the schematic drawing of Fig. \ref{fig:Fig2}(b). Detailed information about the
dielectric hardware and high-pressure equipment can be found elsewhere
\cite{Igarashi,Wence}.

\begin{figure}
	\includegraphics[trim = 10mm 70mm 20mm 70mm, clip, width=0.45\textwidth]{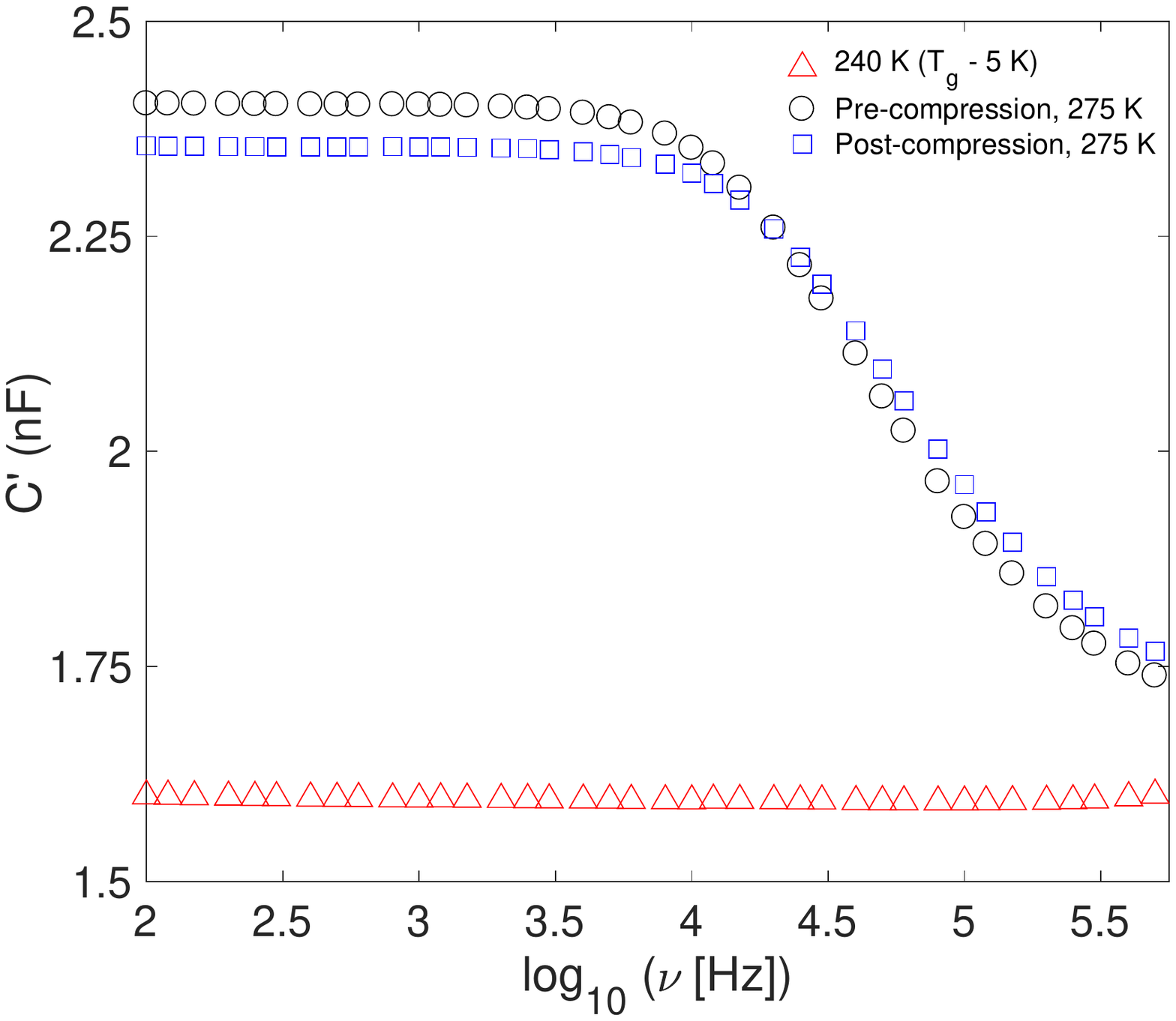}
	\caption{Real part of the capacitance of 5PPE as a function of
		frequency at atmospheric pressure. Spectra at 275 K, before
		(black circles) and after (blue squares) compression at 100
		MPa, show the alpha relaxation. Spectrum at 240 K (red
		triangles) corresponds to the glassy state, conditions at
		which the capacitance at high frequencies is dominated by
		atomic polarization. Data collected with the home-built capacitor presented in Fig. 2(d)}
	\label{fig:Fig3}
\end{figure}

We explored the phase diagram of Fig. \ref{fig:Fig1} by controlling pressure and temperature along the two different pathways. Figure \ref{fig:Fig3} shows the real part of the capacitance for
5PPE at atmospheric pressure at two different temperatures, 275 and
240 K, where the signal is dominated by rotational and atomic
polarization respectively. In order to evaluate the phenomenon of
pressure densification, similar spectra to that shown in
Fig. \ref{fig:Fig3} for 240 K ($T_g$ - 5 K) were analyzed. At 275 K, the signal is dominated by the alpha relaxation of 5PPE and Fig. \ref{fig:Fig3} displays two curves at this temperature, taken before and after compression to 100 MPa. Given the fact that both data at 275 K look similar, it seems plausible that the geometry of the home-built multilayer capacitor is nearly unmodified by compression to 100 MPa. To ensure that rotational polarization does not interfere with our data, the material density was inferred from the capacitance at 240 K and only
the high-frequency flank of the spectra was taken into account.

\begin{figure}
	\includegraphics[trim = 50mm 47mm 15mm 50mm, clip, width=0.7\textwidth]{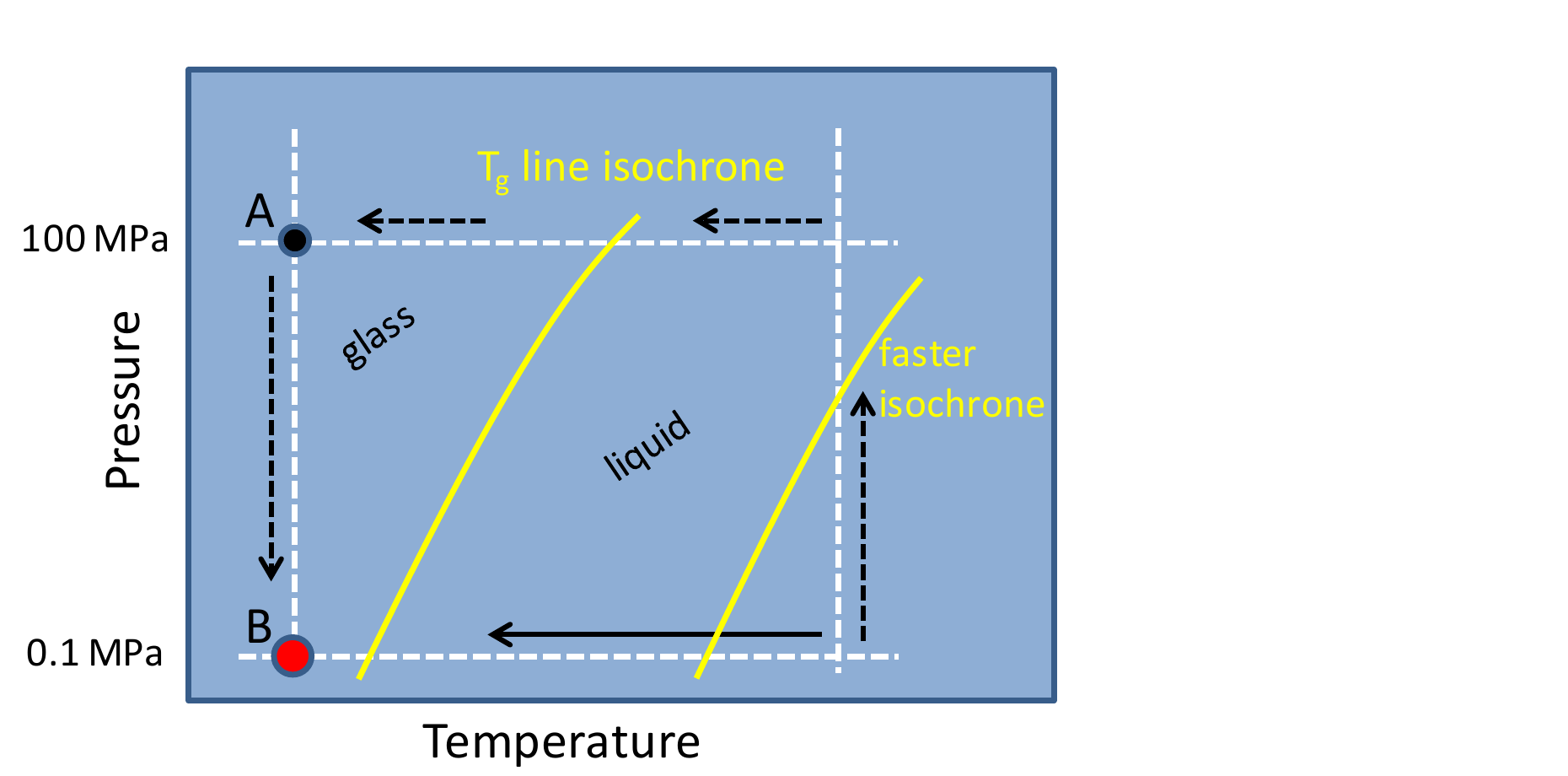}
	\caption{Schematic description of the experimental protocol employed in this study. Two types of glasses were formed by following the solid (cooling at ambient pressure) and dashed (compression-cooling-decompression) black lines, which describe the two different routes for reaching the state point B. Glass and liquid domains are delimited by the $T_{g}$
		line isochrone. A faster isochrone at the high temperature flank is also indicated.}
	\label{fig:Fig2a}
\end{figure}

The protocol used for the experiment is shown in Fig. \ref{fig:Fig2a}. The state-point designated
by the red circle (point B) was reached by the routes indicated by the solid and
dashed black lines. In one case, the sample was cooled at
atmospheric pressure, while in the second scenario the sample was
first compressed at high temperatures, then cooled to the same
temperature (point A), and finally decompressed to 0.1 MPa. We employed the same
cooling rate in both cases. The same thermobaric protocol was applied
on both capacitors.

\begin{figure}
	\includegraphics[trim = 10mm 75mm 15mm 80mm, clip, width=0.5\textwidth]{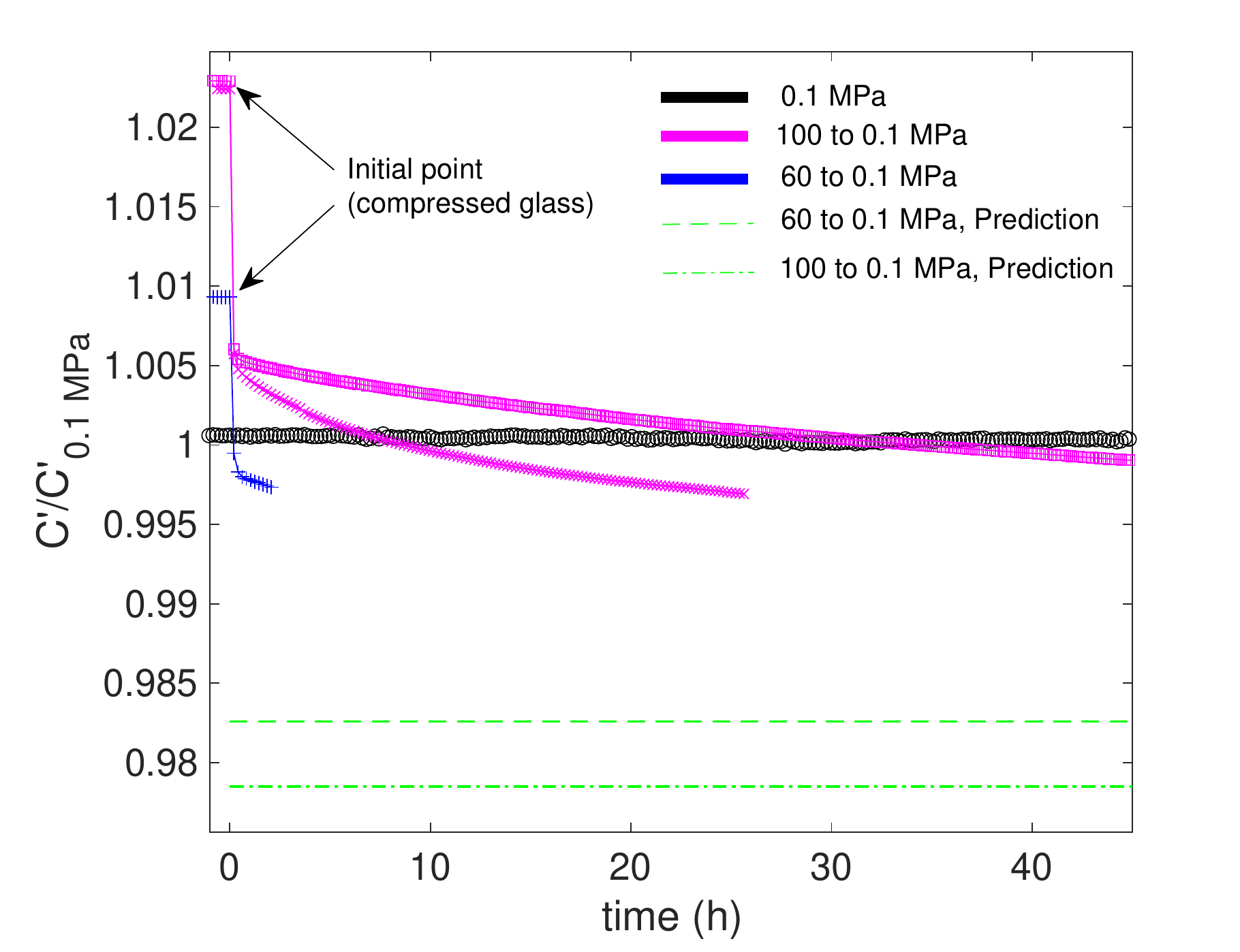}
	\caption{Normalized capacitance at 10 kHz for glasses formed
		by cooling at 0.1 MPa (black circles) and glasses formed by
		cooling at high pressure and subsequent decompression to 0.1
		MPa. Dielectric measurements collected with a multilayer air
		capacitor. Green lines correspond to the value expected
		according to the isomorph theory (absence of pressure densification in R-simple systems) for jumps from 60 and 100 to 0.1
		MPa, together with our assumption that clamped glasses
		cannot flow radially (Eq. \ref{Eq7}).}
	\label{fig:Fig4}
\end{figure}

\begin{figure}
	\includegraphics[trim = 5mm 70mm 15mm 72mm, clip, width=0.50\textwidth]{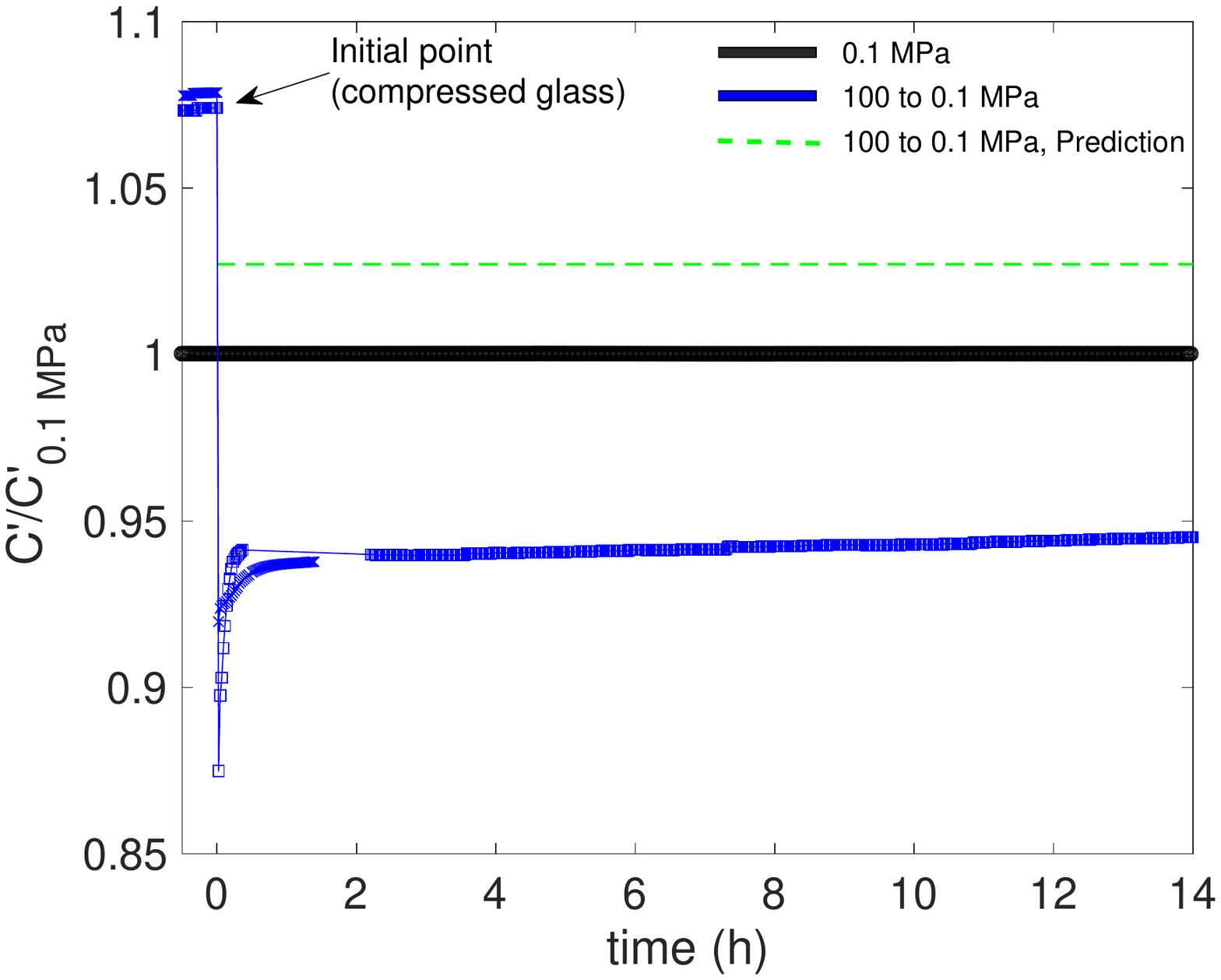}
	\caption{Normalized capacitance at 10 kHz for glasses formed
		by cooling at 0.1 MPa (black) and glasses formed by cooling
		at 100 MPa and subsequent decompression to 0.1 MPa
		(blue). Dashed green line corresponds to the value expected
		according to the isomorph theory (absence of pressure densification in R-simple systems) for a jump from 100 to 0.1
		MPa, together with our assumption that clamped glasses
		cannot flow radially (Eq. \ref{Eq7}). These data were collected using the home-built capacitor
		described in Fig. \ref{fig:Fig2}(d).}
	\label{fig:Fig5}
\end{figure}

\begin{figure*}
	\includegraphics[trim = 0mm 0mm 0mm 0mm, clip, width=0.65\textwidth]{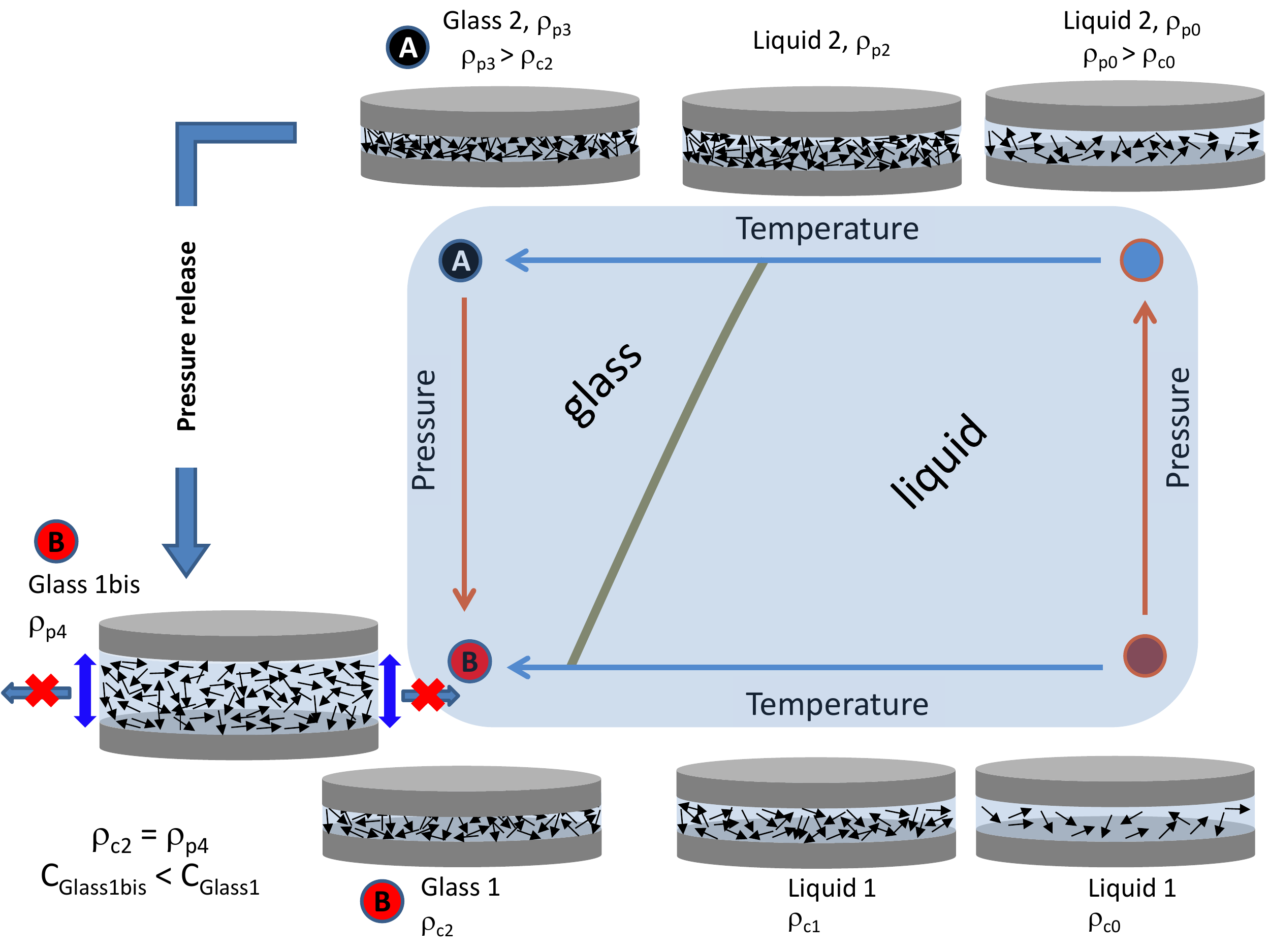}
	\caption{Schematic description of two possible thermobaric routes for reaching the glassy state and the corresponding number density of dipoles (represented by black arrows) in a parallel-plate sample-filled capacitor. The restriction for the glass to flow tangentially after pressure release is indicated in the left bottom corner (Glass 1bis). The pressure release induces expansion in the vertical direction, moving the electrodes apart and consequently reducing the sample capacitance.}
	\label{fig:Fig6}
\end{figure*}

\section{RESULTS FROM THE TWO DIFFERENT CAPACITORS}
Data from the air capacitor shown in
Fig. \ref{fig:Fig2}(c) are presented in Fig. \ref{fig:Fig4}. We report
the real part of the capacitance at a fixed frequency of 10 kHz for
5PPE glasses formed following the paths detailed in
Fig. \ref{fig:Fig2a}. Data are normalized to the capacitance of the
0.1 MPa glass and shown as a function of time. We studied decompression jumps from
100 and 60 MPa to ambient pressure.  Black symbols correspond to a
glass formed upon cooling at 0.1 MPa. When the glass is produced at
the same cooling rate but starting from a liquid compressed at 100
(magenta) or 60 MPa (blue), the capacitance is significantly higher in
comparison to the low-pressure glass, which is expected for a denser
material \cite{Naoki,Kremer}. Then, pressure was released to 0.1
MPa, producing an immediate effect on the capacitive response as
shown in Fig. \ref{fig:Fig4}. The capacitance suddenly decreases upon
decompression and further decreases as a function of time. In the case
of the pressure jump from 60 MPa, the capacitance immediately goes
below the ambient pressure value, and this also happens after 8-30
hours in the case of the jumps from 100 MPa.

In Fig. \ref{fig:Fig5} we show data from the home-built
capacitor with Kapton$\textregistered$ spacers. As stated above, the protocol used
was the same as for the air capacitor. In this case, data are shown for
100-to-0.1 MPa jumps. It is seen that the capacitance of the
post-compressed glass sample immediately goes below the normal glass value. It
makes a fast undershoot and stays stable already after less
than an hour. 

\section{DISCUSSION}
Studying the same sample using the same protocol but in two different
capacitors clearly shows that the results depend not only on the
properties of the glass but also on the sample capacitor. This emphasizes how important it is to
understand the sample geometry when analyzing dielectric data in the
aim of extracting density information. Another important result is
that the capacitance measured after releasing pressure depends on
time, particularly when using the air capacitor, which was also used
in Ref. \cite{Casalini2}, where the authors unfortunately do not report how the the pressure release affects the measured permittivity as a function of time. 

If the glass formed at high pressure has higher ambient-pressure density than the glass
formed at lower pressure, as is the case if the sample
experiences pressure densification, the measured capacitance
is higher in the densified sample. This would be the case
due to the higher number density if the geometry is identical at the
end of the two protocols, as assumed in Ref. \cite{Casalini2}. Moreover, if the sample is compressed in the direction normal to the electrodes, the
effect would be enhanced due to the shorter distance between the
plates.

What we find in Fig. \ref{fig:Fig4} and \ref{fig:Fig5} is not a higher measured capacitance after the high
pressure protocol, but in fact the opposite, a lower capacitance. This can be
understood by the following argument. The glass at high pressure (point A)
has a higher density than the glass at lower pressure (point B). With the
higher density, most scenarios also lead to more dipoles in absolute
numbers between the electrodes of the capacitor than in the low pressure glass. It is for example obvious that
high density leads to a larger number of dipoles in the case where it is
assumed that the capacitor has the same geometry at high and low
pressures. If there is no pressure densification, then the glass formed
at high pressure will attain the density of the low pressure glass
when the pressure is released. However, since it is a glass, it cannot
flow, and it will be clamped in the tangential direction to the
electrode plates. The only way to expand is normal to the electrodes
and this will be done by pushing the electrodes apart (point B in Fig. \ref{fig:Fig6}, Glass 1bis). Consequently,
the final result will be one where the glass formed at high pressure
is in a capacitor with a larger distance between the electrodes and thus
lower empty capacitance, while the actual density and dielectric
constant of the two samples is the same. In the home-built capacitor,
where there is no force keeping the electrodes together, this will
happen abruptly, while it will happen to a lesser extent, and possible over time, in
the air capacitor where it requires bending of the electrodes. The
slow response seen in the air capacitor could also be due to actual
flow taking place. At 240 K, the alpha relaxation time for 5PPE is estimated to be $9.2\times10^{3}$ s and
with a distance between the plates of 0.2 mm (air capacitor), this gives a characteristic flow-time of $2.3\times10^{7}$ s, following the argument in Ref. \cite{Kristine2}. This flow time is long compared to the time scale of our experiment.

Considering the case of the home-built capacitor, we can try to make a
quantitative prediction of what a pure longitudinal expansion would
give, starting from the value of the capacitance measured at high
pressure. Assuming that the glass does not expand tangentially to the
electrodes as the pressure drops, i.e., that it can only expands along the normal direction, the derivative of \textit{$C_m$}
with respect to pressure following Eq. (\ref{Eq2}) is given by
\begin{equation}\label{Eq4}
\frac{dC_m}{dp}=\varepsilon_\infty \frac{dC_g}{dp}+C_g\frac{d\varepsilon_\infty}{dp}.
\end{equation}
Given that the longitudinal isothermal
compressibility is defined as 
\begin{equation}\label{Eq5}
\kappa_T=-\frac{1}{l}\left(\frac{\partial l}{\partial p}\right)_T,
\end{equation}
where \textit{l} corresponds to the distance between the electrodes (dimension in the direction of the flow), Eq. (\ref{Eq4}) can be written as \cite{Kristine2} 
\begin{equation}\label{Eq6}
\frac{1}{C_m}\frac{dC_m}{dp}=-\frac{\varepsilon_\infty+B(\varepsilon_\infty)}{\varepsilon_\infty}\kappa_T,
\end{equation}
with $B(\varepsilon_\infty)=\frac{(\varepsilon_\infty-1)(\varepsilon_\infty+2)}{3}$.
\\

At 0.1 MPa and 240 K, $\varepsilon_\infty$ for 5PPE takes a value of 3.4 approximately. Thus, Eq. (\ref{Eq6}) can be reduced to
\begin{equation}\label{Eq7}
\frac{1}{C_m}\frac{dC_m}{dp}\simeq-\frac{2.3}{M},
\end{equation}
where \textit{M} is the isothermal longitudinal modulus which is
defined as the inverse of the longitudinal isothermal compressibility \cite{Kittel}. The
longitudinal modulus is given by the bulk, $K$, and shear, $G$, moduli
for isotropic solids through the equation $M=K+4/3G$. From the
longitudinal modulus $M$ and, by means of Eq. (\ref{Eq7}), we can make a
prediction of the measured capacitance for a 5PPE glass after
releasing pressure. This is possible thanks to the availability of data for $K$ and $G$ for 5PPE at temperatures close to 240 K
\cite{Tinathesis}.

The decrease in capacitance calculated from this procedure for the home-built capacitor is shown as a green dashed line in Fig. \ref{fig:Fig5}. Curiously, the measured decrease in
capacitance is much stronger, than what we get from this
calculation. This could be due to cracking of the sample as a
consequence of the abrupt change in pressure. This would lead to an
even larger distance between the electrodes than a simple expansion. The resulting capacitance after decompression predicted by Eq. \ref{Eq7} is also shown in Fig. \ref{fig:Fig4} for the air capacitor.

In Fig. \ref{fig:Fig6} we show a schematic description of the situation explained above that allows one to understand why the post-compression glass presents a lower capacitance than the normal one. The number density of dipoles between the electrodes as a function of temperature and pressure is illustrated for the normal (Glass 1), compressed (Glass 2) and post-compression (Glass 1bis) glasses. The existence of isomorphs in strongly correlating liquids gives a theoretical foundation for arguing that, on cooling, both normal and compressed glasses of 5PPE explore state points in the phase diagram that remain $\textit{quasi-isomorphic}$ \cite{Gnan,Fragiadakis}. Therefore, post-compression and normal glass should have similar densities that, together with the longitudinal expansion of the glass in the normal direction, would give rise to a lower value of the capacitance for the post-compression glass as indicated at the bottom of Fig. \ref{fig:Fig6} (Glass 1bis, point B).  

\section{CONCLUSIONS}
Our data do not match the results reported by Casalini
\cite{Casalini2} on DC704 and do not indicate any significant
pressure densification effects in 5PPE. While a signature of pressure
densification would have given rise the capacitance to values higher
than those shown by the 0.1 MPa glass, we observe here that the
post-compression glasses present even lower values. In the case of the air
capacitor we only see the lower values after significant waiting
time. The discrepancy between the results could be explained if the
data shown in Ref. \cite{Casalini2} were measured immediately after the pressure was released.

We cannot determine on whether or not simple glass formers
such as 5PPE can be pressure densified, but this investigation has highlighted the challenge of utilizing dielectric spectroscopy to tackle this problem
effectively. The measured capacitance depends not only on the
dielectric constant, but also on the geometry of the capacitor. In
order to convert the measurements to density it is therefore
crucial to understand the geometry. By comparing the results from two
different capacitors with the same sample and protocol, we have demonstrated
the importance of the geometry. Moreover, it should be stressed that a glass cannot be assumed to flow until waiting times are much longer than
the alpha relaxation time. Density changes when releasing pressure
in the glassy state therefore must take place by a solid-like type
of deformation, which involves distortion of the dielectric cell.

\begin{acknowledgments}
This work is supported by the VILLUM Foundation grant Matter (16515).
\end{acknowledgments}

\end{document}